# The Very Big ILC

David G. Hitlin

California Institute of Technology

Pasadena, CA 91125 USA

September 1, 2020

## Abstract

In the spirit of Leon Lederman's 1977 proposal for the siting of the VBA, we propose a version of the International Linear Collider along the US-Mexico border

As the controversy over President Trump's border wall has once again come to the fore, it is worth considering an idea that could put the project on a more positive trajectory. This concept could provide the desired deterrence capability without reallocation of Defense Department funds, the use of alligators, the separation of children and parents, or the need to shoot anyone in the leg.

President Trump's Wall has been controversial, the cause of an extended partial government shutdown and of a national emergency declaration, in part because of its associations with medieval structural fortifications. Could the functionality of the Wall be retained in a more modern context? I believe this is possible, with great benefit to elementary particle physics as a corollary. Hence the modest proposal presented herein, to build a super version of the long-sought International Linear Collider (ILC) along the US-Mexico border.

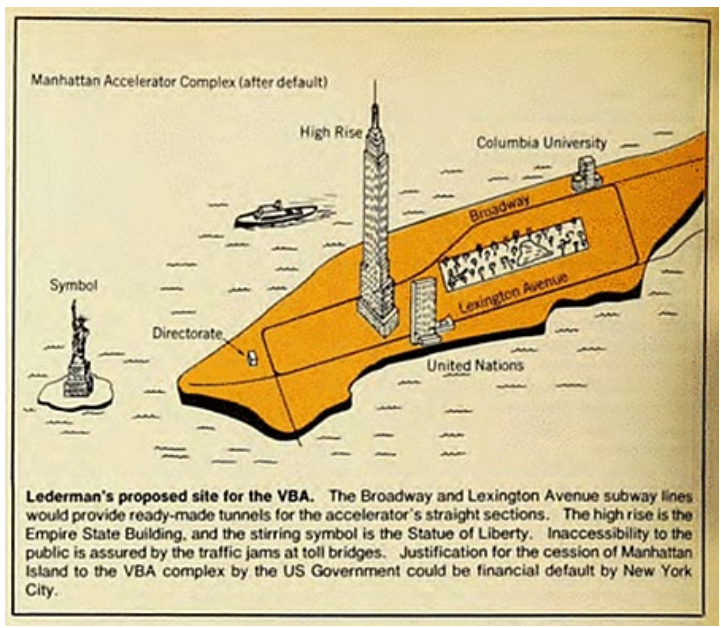

**Figure 1**. Leon Lederman's proposal in Physics Today to site the VBA in the Manhattan subway tunnel system.

There is an antecedent to this type of creative accelerator siting idea. In 1977 Leon Lederman, a prominent proponent of what was then called the Very Big Accelerator (VBA) [1], advocated in Physics Today [2] building the VBA, an SSC or LHC-scale collider, in the New York City subway tunnel system (see Figure 1). This proposal was made, as is the current one, during a crisis, one engendered by a subway workers' strike and the desperate financial straits of the New York City government [3], hence Leon's reference to "financial default", which would have made the tunnels available for this worthy new purpose.

The length of the US-Mexico border from Tijuana to El Paso is about 1160 km (see Figure 2). There is a region near Nogales, AZ (red circle) that provides a natural site for the interaction region of the VBILC, perhaps better named the TrumpILC, by analogy with the naming of the SSC Laboratory after Ronald Reagan. The border extends in a straight line westward from Nogales for 376 km and eastward for 273 km. The intersection is at a 21-degree angle, somewhat larger than the 14 mrad crossing angle of the ILC, but modification of the interaction region to accommodate a larger angle may be feasible. If it is not, the region from south of Yuma to Nogales provides an entirely straight segment of comparable length as well as greatly improved beach access.

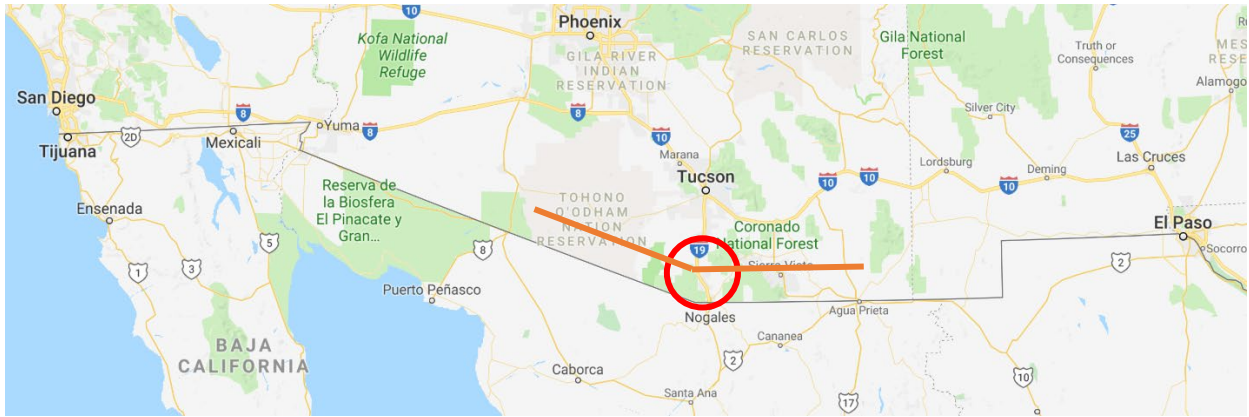

**Figure 2.** The US-Mexico border from Tijuana to El Paso (from Google Maps). The position of the 300 km long TrumpILC, with its interaction region near Nogales, AZ is shown in orange.

Modern high-energy accelerators are typically housed in tunnels to provide the needed radiation shielding, as is the case with the various versions of the ILC considered, for example, in the 2013 Technical Design Report [4]. Depending on the particular site, the tunnel depths considered are as shallow as 10 meters, which is still more shielding than is actually required. A tunnel along the US-Mexico border would not, however, serve the dual-purpose intended here, and would, in addition, likely interfere with the existing network of cross-border tunnels (at what depth do eminent domain laws cease to apply?). A recent study [5] of the amount of shielding needed between the two sections of the ILC tunnel concludes that 2.5 meters of concrete is sufficient to reduce the external radiation to 25 µSv/hr. Thus, one can contemplate building the TrumpILC on the surface, with concrete shielding blocks of this thickness serving the dual purpose of border protection and (perhaps asymmetric) radiation shielding.

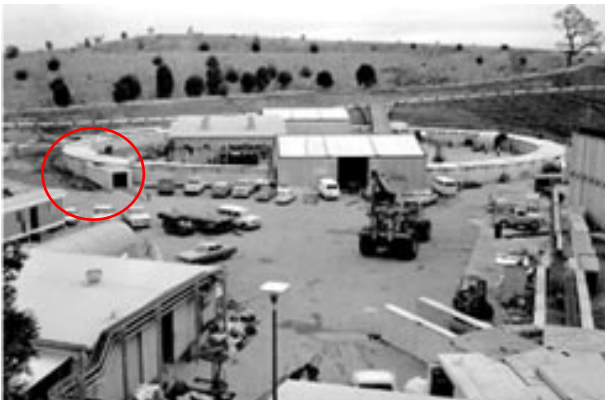

There is relevant precedent here as well. The SPEAR $e^+e^-$ collider was not housed in a tunnel; it was located above ground in a parking lot in the SLAC Research Yard, with radiation shielding provided by freestanding concrete blocks (see Figure 3). The dark square aperture visible on the left side, part of the then unfinished injection line, shows the thickness of the radiation shielding required for a beam energy of up to 3.6 GeV.

**Figure 3.** The SPEAR ring under construction at SLAC.

Our proposal is thus to build the TrumpILC on the surface, along the US-Mexico border, centered near Nogales, AZ, with a simple concrete structure providing both radiation shielding and the border wall. You can't get any less medieval. Such a structure, built of prefabricated concrete blocks, would be far less expensive than a tunnel of equal length. It is possible, of course, to improve the deterrent function of the Wall by reducing the thickness of the shielding on the Mexican side of the structure. A simple value-engineering pass can optimize the thickness of the southern shielding wall, using applicable criteria provided by the Department of Homeland Security.

Note that the TrumpILC would manifestly be an *International* Linear Collider. This designation could be further enhanced in two ways: the structure could be built with the accelerator on the border itself, so that the shielding structure encroaches on Mexican territory, or we could actually have Mexico pay for the TrumpILC or a portion thereof. The Mexican government would likely be more willing to fund this structure as part of an international scientific consortium than to pay for the current single-function wall.

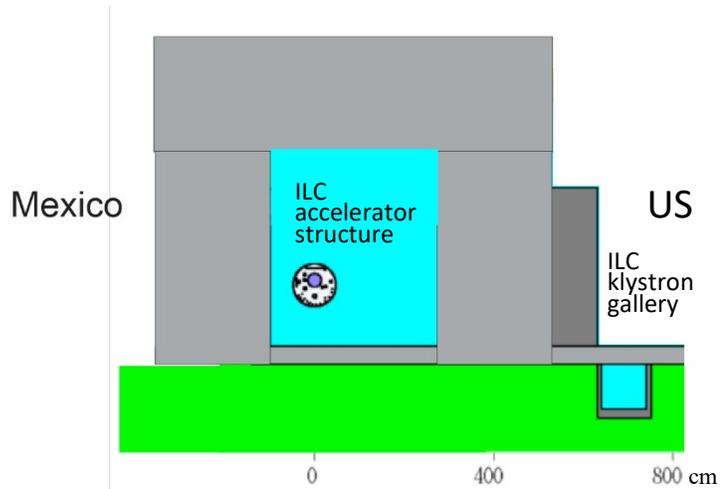

**Figure 4**. Cross section of the TrumpILC accelerator housing

Figure 4 show a cross section of the TrumpILC accelerator housing. With the accelerator structure on the surface, the klystron gallery and other associated equipment does not have to be in a shielded housing, but can be located, as with the original SLAC two-mile accelerator, in unshielded above ground structures located on the US side. A few comments on the structure itself: The President's latest notion of a "beautiful steel slat" wall is clearly not compatible with our shielding requirements; it would be necessary for him to backtrack to his original concept of a concrete wall; a minor concession in this context. The shielding requirements also do not extend quite to the height of Trump's Wall specification. This could easily be dealt with by a simple razor wire fence on top of the shielding (the National Guard is already on site, and has demonstrated its capabilities in razor wire deployment), by installing shorter beautiful steel slats at the top, or even by leaving a small vertical crack in the shielding, with appropriate radiation hazard signage.

Is the ground in the desert sufficiently stable for ILC collisions? As there is little traffic noise, the answer is likely yes, provided that the number of people attempting to scale the structure at any given time can be kept to a minimum (see above). It may also be necessary to deal with the Second Amendment-based problem faced by LIGO Livingston; the northern wall of the klystron gallery walls may have to incorporate a Kevlar layer.

The ILC under consideration in Japan is 31 km long, with an initial center-of-mass energy of 250 GeV, extensible to 500 GeV. We could, of course, just replicate this design for the Nogales site, but this would be an opportunity missed. The length of the US-Mexican border that we could potentially protect provides us with the ability to greatly expand the reach of the TrumpILC. We therefore propose a design that is 300 km long (150 km for each arm), which could have a center-of-mass energy of 5 TeV, assuming the use of the same 31.5 MeV/m superconducting RF cavities used in the ILC. The astute reader will recognize that a linear accelerator of this length on the surface must confront the issue of the curvature of the earth. As many Trump supporters do not recognize this as a concern, we propose to leave the solution of this problem to them. This could provide, in the words of former President Obama, "a teachable moment".

If we take the President's initial demand of $5.7B for his border security wall as a constraint, the starting center-of-mass energy of the TrumpILC could be scaled to fit this budget, while still preserving the ultimate 5 TeV capability of the collider. A more capable design with each arm as long as 270 km is also possible, reaching a center-of-mass energy of about 8 TeV, but the HEP community long ago recognized the wisdom of moderating the size of its requests for research facility funding.

The proposed TrumpILC is, of course, a very large structure. The Grand Coulee Dam in Washington contains a bit more than 9 million cubic meters of concrete. The full 300 km TrumpILC, together with the associated damping rings and experimental detector housings, would require less than twice this amount, and this could be reduced by grading the shielding along the length of the accelerator, as well as by enhancing the deterrent factor of the structure on the Mexican side. The President has been clear about the people he wants his Wall to keep out: "They're bringing drugs. They're bringing crime. They're rapists. And some, I assume, are good people" [6]. We could certainly save a substantial volume of concrete by reducing the thickness of the top, as well as the southern side, shielding blocks. If just the top blocks were thinned, the people exposed to higher levels of radiation would be only those undesirables who made it to the top of the TrumpILC shielding and through the beautiful steel slats or razor wire. We could easily reduce the size of this endangered cohort by administrative means, by for example, posting bilingual signs directing murderers, rapists, drug mules, and even good people, to the nearest conventional port of entry.

**Disclaimer**

This note is my personal contribution to the ongoing conversation concerning a border wall, making use of my professional expertise and experience. The content does not represent the view of California Institute of Technology or of the Federal agencies that have funded my research in elementary particle physics.

This drafting of this note was begun early in the Trump administration, as the ILC approval process ground on. I have circulated versions to friends involved in the ILC effort. Their response was typically that this document, while amusing, "would not be helpful". As a supporter of the ILC, I therefore refrained from broader distribution. There should, however, be a statute of limitations on the suppression of candid discourse, which I would contend has been far exceeded by the halting and protracted Japanese government decision process. Given the renewed topicality of the Wall in the context of the 2020 presidential election, I thought it timely that the note see the light of day.

## Acknowledgements

I thank J.-F. Caron for pointing out that the earth is not flat (in his opinion).